# Builders Instead of Consumers: Training Astronomers in Instrumentation & Observation

Sarah Tuttle, Hanshin Lee, Cynthia Froning, Mike Montgomery for McDonald Observatory

The astronomy community has made clear our shared scientific vision in the Astro2010 decadal survey. Who will build this future? The cost and scarcity of telescope resources makes vital "learning through doing" extremely difficult for students and early career researchers. What is needed now and in the future to provide a depth of knowledge, creativity, and experience in our field? At McDonald Observatory we have a clear model in answer to that question, and a long history of successfully training the next generation of instrument builders and observers. We must support and sustain small to medium range "local" resources such as McDonald to foster the successful growth of our field.

McDonald Observatory offers a unique example of a holistic model to train young astronomers, provide a test bed for novel instrumentation, and use smaller telescopes for exciting science. In this age of increasing project scale in ground-based O/IR it is crucial to retain telescopes to provide opportunities to new astronomers. As NWNH and the portfolio review calls out specifically – we must maximize science, build on current endeavors, and balance current and future programs. With decreased financial resources, each investment must meet multiple needs. We believe our model not only meets but exceeds that requirement.

The facilities at McDonald Observatory provide a powerful platform for training and mentorship of undergraduate students, graduate students, and postdoctoral researchers. Students have the opportunity to undertake observing research programs that can be led by the students themselves at every stage. These observing programs are often innovative or experimental projects that new researchers find unsupported on highly subscribed facilities. Similarly, McDonald Observatory has historically played host to numerous instrumentation projects that give enormous latitude for student participation throughout the project. We provide here a small representative set of recent examples.

**Graduate & Postdoc Led Instrumentation:**

The current upgrade of the HET for HETDEX has featured significant leadership roles for two postdoctoral researchers both on the telescope side (Wide Field Upgrade, Lee) and the instrument side (VIRUS, Tuttle 2014). This is possible because McDonald has an ethos of encouraging early career researchers and empowering them in leadership positions.

All telescopes on the mountain are fair game. A student completely reworked the 30" telescope to develop prime focus camera imaging capability in pursuit of his dissertation in 1992 (Claver). Students and post-docs were involved in sub-system development for the HET that persists to this day, working on mirror alignment systems (MARS) and the segment alignment/maintenance system (SAMS) (Wolf 2003A & B, Palunas 2004). Optical design for the HRS upgrade was led by a postdoctoral researcher (Barnes).

The 2.7m is a very fruitful development platform. Three of the current high subscription instruments (Mitchell Spectrograph (VIRUS-P), IGRINS, and 2dcoude) made heavy use of postdoctoral and graduate student work. The instruments also demonstrate the broad range of technology development being done – from novel IFU instrumentation, to high resolution infrared spectroscopy with immersion gratings.

Technology development is also led by graduate students. A volume phase holographic (VPH) grating test bench was built and used by several projects, including VIRUS-P and IGRINS (Adams 2008). A modified version was developed and built to support VIRUS grating manufacture (Chonis 2012). Fiber testing, including test bench development, lifetime testing, and manufacturing protocol was led by a graduate student (Murphy 2013).

**Graduate Led Observing Programs**:

In combination with the University of Texas at Austin astronomy department, McDonald Observatory encourages and supports student-PI projects.
Some work ties in directly to large project development. The HETDEX Pilot survey used VIRUS-P (a test instrument that became a facility instrument) to validate the feasibility of the HETDEX survey method (Adams 2011). The students leading the survey also developed VACCINE, the data reduction software package to support the instrument.
A compelling example of graduate student led projects is the VIRUS-P Exploration of Nearby Galaxies on the 2.7m telescope (VENGA; Blanc 2013). Students gained invaluable experience designing the survey, conducting the observations, and developing the data reduction pipeline. The 2.1m was host to a recent exciting and epic dissertation studying extremely low mass white dwarfs over 220 nights (Hermes 2013).

**Classwork and Undergraduate Involvement**:

The Freshman Research Initiative (FRI) program at UT is a college-wide program in Natural Sciences for directly involving freshman in faculty research. UT offers two streams, involving over 40 freshman a year in cutting edge astronomy research. The highlight of their experience involves trips to McDonald Observatory to take data directly related to their research projects, as well as to assist with on-site public star parties. These research and teaching experiences are formative for the students, and begin their transformation from student to scientist.

Our instrumentation class is taught by a rotation of our instrumentally inclined faculty. Students join from engineering and physics disciplines, and the class includes graduate and undergraduate students. The experience is invaluable, and we often discover students who join observatory based research projects. Replicating VIRUS utilizes several undergraduates in assembly roles, some discovered through this class.

**Observatory Careers:**

Given the wide array of projects in observing and instrumentation available to students at McDonald Observatory, it is no surprise that alumni of the UT Austin Astronomy undergraduate and graduate programs are well represented in a variety of positions in research astronomy. Alumni have moved into positions that include faculty appointments but go beyond these to encompass roles in observatory management and operations (at Keck, NOAO, Gemini Observatory, Space Telescope Science Institute, among others), instrument and telescope development (including HST, LSST, LCOGT, and SOFIA), teaching, and public outreach. It is this success that should motivate us. The projects the community is invested in must be created and maintained, and that training requires more than a "data consumer" experience.

We must address the issue of training, and how that integrates into the larger question of facilities and funding. We believe we can retain telescopes for training and technology development while getting high quality science out. It is crucial to retain capacity and opportunities for students to develop technical and leadership skills. McDonald observatory has demonstrated one successful model that should be emulated and supported moving forward.